\documentclass[aps,manuscript]{revtex}
\usepackage[T1]{fontenc}
\usepackage[latin1]{inputenc}
\usepackage{graphics}

\makeatletter

\providecommand{\LyX}{L\kern-.1667em\lower.25em\hbox{Y}\kern-.125emX\@}

\makeatother

\begin{document}

\title{
\begin{flushright}
{\rm UPR-873T  \\
CERN-TH/2000-018}
\end{flushright}
A Cosmological Mechanism for Stabilizing Moduli}

\author{Greg Huey$^1$, Paul J. Steinhardt$^1$, Burt A. Ovrut$^2$,
and Daniel Waldram$^{1,3}$}
\address{$^1$Department of Physics,
Princeton University \\
Princeton, NJ 08544-0708,  USA}
\address{$^2$Department of Physics and Astronomy,
University of Pennsylvania \\
Philadelphia, Pennsylvania 19104-6395, USA}
\address{$^3$Theory Division, CERN CH-1211, Geneva 23, Switzerland}

\maketitle
\begin{abstract}
In this paper, we show how the generic coupling of moduli to the 
kinetic energy of ordinary matter fields results in a 
cosmological mechanism that influences the evolution and stability
of moduli.  As an example, 
we reconsider the problem of stabilizing the dilaton 
in a non-perturbative potential
induced by gaugino condensates. A well-known difficulty is that the potential
is so steep that the dilaton field  
tends to overrun the correct minimum and to evolve
to an observationally unacceptable vacuum.
We show that the dilaton coupling to the kinetic or
thermal energy of matter fields 
produces a natural mechanism for
gently relaxing the dilaton field into the correct minimum of the potential
without fine-tuning of initial conditions. 
The same mechanism is potentially relevant for stabilizing other moduli fields.
\end{abstract}

A fundamental problem in supergravity and superstring theories is the stabilization
of moduli fields, particularly the dilaton. Perturbatively, $\Phi\equiv
{\rm exp}(\lambda \phi)$ (the dilaton) has
no potential, although it does not behave as a free field because it has non-linear
couplings to the kinetic energy of the axion field. 
(Throughout this paper, we use $\Phi$ and $\phi$ interchangeably to 
represent the dilaton according
to convenience; the  constant $\lambda = \sqrt{16 \pi}/m_{pl}$, where 
$m_{pl}\equiv 1.2 \times 10^{19}$~GeV is the 
Planck scale, is chosen so that
$\phi$ has a canonical kinetic energy density, $\frac{1}{2} \dot{\phi}^2$.)
A non-perturbative potential
can be induced by gaugino condensates \cite{GCPot_1,GCPot_2,GCPot_3}. With several
gaugino condensates, parameters can be tuned so that there is a locally stable
minimum with zero cosmological constant \cite{3GCPot}. See the solid curve in
Fig.~1. However, the potential
is exponentially steep ($V\sim \exp (-\exp (\phi )) $) and the desired
minimum, $\Phi_{min}$,
is separated by an exponentially small barrier (compared to the Planck scale)
from an observationally unacceptable anti-de~Sitter vacuum \cite{Chall_SSC}.
It appears that, unless the initial conditions of the dilaton field are finely-tuned
to lie very near the correct minimum, the field will overrun or miss altogether
the desired minimum. 

In this paper, we present a possible robust solution to this problem based on
generic properties of the dilaton and natural cosmological effects. 
The solution relies on the
coupling of the dilaton to the kinetic energy density
of ordinary matter fields which has important consequences
in the early universe when the 
thermal (kinetic)  energy density is high. In the radiation-dominated epoch,
at least three  effects come into play, two of which have been
considered previously.

First, the energy density in the thermal component increases the Hubble damping,
as emphasized by Barreiro \emph{et al} \cite{Track_racetrack}. If the thermal
energy density is very large compared to the dilaton energy density, the Hubble
damping factor is significantly enhanced and the evolution of the dilaton is
slowed. As a result, $\Phi$ can be allowed somewhat smaller initial values
(corresponding to climbing further up the steep part of the potential in
Fig.~1)  and still be trapped at $\Phi_{min}$. This
is  a modest expansion in allowed initial conditions.  In the scheme
presented here, we find that the range of allowed initial conditions is 
enormously expanded.

Second, as pointed out by Horne and Moore \cite{ChaotCouplConst}, the dilaton
couples non-linearly to the axion field and, if both fields have large initial
kinetic energy densities compared to their potential, the non-linear coupling 
causes  $\Phi$ to undergo chaotic motion back and forth in its potential over
a finite range in $\Phi$ that  includes the desired minimum. 
If the chaotic behavior
could be sustained, then this would
enhance the probability that \( \Phi  \) is trapped in the correct minimum.
However, as pointed out by Banks \emph{et al} \cite{ModulCosmo}, the axion
kinetic energy decays too quickly and spatial inhomogeneities grow too rapidly
during the chaotic phase.

This paper points out a third feature of the dilaton in a cosmological setting
that can
provide a robust mechanism for dilaton stabilization. Namely, although
the dilaton couples non-perturbatively to itself, it 
couples \emph{perturbatively}
to the kinetic energy and potential energy 
of all matter and gauge fields. In studying vacuum solutions,
these fields and their kinetic energies are usually set to zero. However, in
a cosmological setting, they produce a non-negligible, temperature-dependent
contribution to the dilaton effective potential that can allow the dilaton field
to be gently lowered into the desired minimum as the universe expands and cools.
Whether this mechanism works depends on the functional form of the
dilaton coupling to the matter and radiation energy densities. If we take forms
suggested by superstring theory, the scenario works. 
(When the first two effects above, Hubble damping and coupling to the axion,
are also included, they help to
extend the range of dilaton couplings which work.)

We write the lowest component of
dilaton superfield as \( S=\Phi +iA/m_{pl} \), where \( \Phi  \) describes
the dilaton and \( A \) the axion. 
The non-perturbative dilaton potential, $V_{np}$, is due to multiple gaugino
condensates, arranged to yield a stable minimum with zero cosmological constant
($\Phi =\Phi_{min}$): the racetrack model \cite{3GCPot} as shown
 in Fig.~1.
The energy scale has been blown up by more than $60$ orders of magnitude
compared to the Planck scale in order to make visible the features near $\Phi_{min}$.
The minimum is locally stable.
 There is a barrier at \( \Phi >\Phi _{min} \)
peaking at $\Phi =\Phi_{p}$ which separates the desired minimum from an
anti-de~Sitter vacuum. 
The height of the barrier is tiny, typically 50 or more orders
of magnitude below the Planck density.
At $\Phi <\Phi_{min}$ the potential rises exponentially
steeply to values $V_{np}[\Phi ]\gg V_{np}[\Phi_{p}]$. 

Based on this description and Fig.~1, it is simple to see why it is hard
to be trapped at $\Phi=\Phi_{min}$.
If $\Phi$ begins at $\Phi_0 >\Phi_{p}$, on the right side of the barrier
from $\Phi_{min}$, it is unlikely to be trapped at $\Phi_{min}$.
For $\Phi_{0}<\Phi_{min}$, there is a very limited range 
of initial conditions for which $\Phi$
is trapped at $\Phi _{min}$. In particular,
 if $V_{np}[\Phi_{0}]\gg V_{np}[\Phi_{p}]$,
(e.g. if the initial potential energy density is near the Planck scale
or compactification scale, which is much greater than the barrier height)
the field tends to roll rapidly down
the exponential potential, overshooting $\Phi_{min}$ and
the barrier (\( \Phi =\Phi_{p} \)), ending up in the 
wrong vacuum.

At high temperatures
the relevant terms  of a typical Lagrangian have the form: 
\begin{equation}
\label{4DeffL_gen}
\sqrt{\left| g\right| }L=\sqrt{\left| g\right| }\left\{ \frac{1}{2}\left( \partial \phi \right) ^{2}+
\frac{f_A(\Phi)}{2}\left( \partial A\right) ^{2}
+\frac{f(\Phi )}{2} | \partial C|^{2} - g(\Phi) V_C(C) -V_{np}(\Phi, A)\right\} 
\end{equation}
where 
\( C \)
is the complex scalar field in a chiral supermultiplet (a matter field) 
with potential $V_C(C)$,
$f_A(\Phi)\equiv 1/2 \Phi^2$ is the dilaton-axion coupling, 
and
\( f(\Phi ) \) and $g(\Phi)$ are, respectively,
the coupling of the dilaton to the 
kinetic energy and potential energy of $C$.
The exact form of $f(\Phi)$ and $g(\Phi)$
depends on the theory one is considering
(see below).
\( V_{np}(\Phi, A) \) is the racetrack potential,
constructed from the superpotential
\begin{equation}
\label{3gc_SupPot}
W\propto m_{pl}^3 Z(Z+1)^{2}\; \; \; \; ;\; \; \; Z\equiv e^{-\alpha S}
\end{equation}
and K\"ahler potential
\begin{equation}
\label{3gc_KahlPot}
K=-m^{2}_{pl}\ln \left( S+\overline{S}\right) -\ldots.
\end{equation}
Here $\alpha$ is a constant whose value depends on the gauge group.
The result for the potential is 
\begin{equation}
\label{3cg_Vnp}
V_{np}=e^{K/m^{2}_{pl}}\left[ K^{S\overline{S}}D_{S}W\overline{D_{S}W}-\frac{3}{m^{2}_{pl}}W\overline{W}\right] =\frac{1}{\Phi} \sum ^{5}_{j=1}h_{j}(\Phi ,A)e^{-(j+1)\alpha \Phi }
\end{equation}
 here \( D_{S}W\equiv \partial _{S}W-K_{S}W/m^{2}_{pl} \) and the \( h_{j}(\Phi ,A) \)
are polynomials of degree \( 2 \) in \( \Phi  \). The functional form of \( W \)
 is  chosen such that the cosmological constant is zero at
the minimum.
From Eq.~(\ref{3cg_Vnp}) we can see that \( \Phi  \) decreases exponentially
fast for \( \Phi <\Phi _{min} \); and, as proven in \cite{Chall_SSC}, using
the holomorphic property of \( W \), \( V_{np} \) is forced to have a barrier
at some \( \Phi =\Phi _{p}>\Phi _{min} \) separating \( \Phi _{min} \) from
an anti-de~Sitter minimum at \( \Phi >\Phi _{p} \). See Fig.~1.

\begin{figure}
{\par\centering \resizebox*{4in}{4in}{\includegraphics{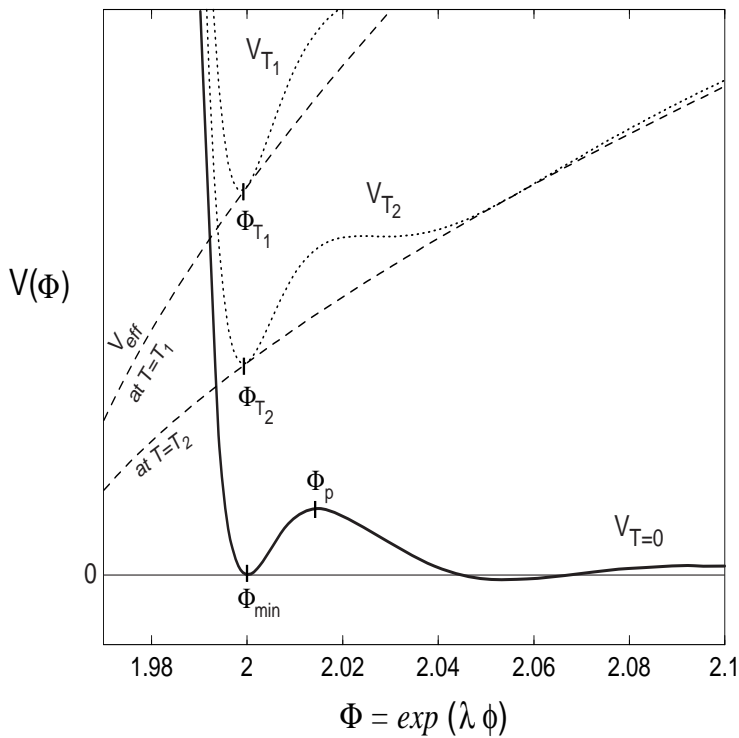}} \par}
\caption{\label{fig_StabPot} A schematic of the
racetrack  potential for the
dilaton $\Phi= {\rm exp}( \lambda \phi)$, 
generated by gaugino condensates ($\lambda$ is a
constant). This is represented by the solid curve.
The desired minimum
at $\Phi=\Phi_{min}$ is
separated by a small barrier, peaked at $\Phi=\Phi_p$. Beyond  
 $\Phi=\Phi_p$ (around $\Phi= 2.05$ in this example), there is an
 unacceptable anti-de Sitter vacuum.
(The energy
scale has been blown up by more than $60$ orders of magnitude to make
the barrier visible.) The dashed line represents 
$V_{eff}$, the effective potential for $\Phi$
 stemming from
the dilaton coupling $f(\Phi)= g(\Phi) = 1/\Phi$
at temperature $T=T_i$.
As $T$ decreases from $T_1$ to $T_2$ to zero, this contribution
adiabatically decreases. The dotted line represents the total  finite 
temperature potential for $\Phi$,
 $V_{T_i}$, which has a minimum at $\Phi=\Phi_{T_i}$.}
\end{figure}

Note that Eq. (\ref{4DeffL_gen}) includes a perturbative coupling of $\phi$
to the kinetic energy of the $C$ field. In previous treatments of 
dilaton stabilization at the minimum of
racetrack potentials, this coupling was 
ignored because the kinetic energy was treated as negligible. While this is 
justified at zero temperature, the kinetic energy is non-negligible
at high temperature and, then, this dilaton coupling is extremely 
important and should not be ignored.

Stabilization can result  under two conditions: (a) coherent oscillation
of a homogeneous scalar (matter) field; and (b) thermal excitation of 
matter fields.  Both are plausible sources in the early universe.
Let us first consider Case (a), the coherent oscillations of a scalar
field $C$.   If the potential energy is $V_C \propto |C|^n$ for 
integer $n \ge 2$, then the 
oscillatory 
$C$-field energy density $\rho_C$ decays as $a^{-6n/(n+2)}$. For 
simplicity, we will restrict ourselves to $n=4$ for which 
$\rho_C \propto a^{-4}$, similar to  radiation. Furthermore, 
we take $f(\Phi)=g(\Phi)$.
Because the field is assumed to be homogeneous, $\nabla C = 0$.
Then, the action in 
 Eq.~(\ref{4DeffL_gen}) contains the interaction
 $f(\Phi) \left[\frac{1}{2}|\dot{C}|^2- V_C(C) \right] \equiv f(\Phi) p_C$,
where $p_C$ is the pressure of the oscillatory scalar field.
Assuming a Friedmann-Robertson-Walker metric,
the equation of motion for $\Phi= {\rm exp}(\lambda \phi)$ becomes
\begin{equation}
\label{eq3}
\frac{1}{a^{3}}\frac{d}{dt}\left( a^{3}\dot{\phi }\right) \, -\, f'
 p_{C} +V_{np,\phi }=0
\end{equation}
where $a(t)$ is the Robertson-Walker scale factor and $f' = df/d\phi$.
According to Eq.~(\ref{eq3}), the pressure due to $C$ exerts a
force on $\phi$ equal to $- f' p_C$.  
From the equation-of-motion for $C$, we see
\begin{equation}
\label{eq2}
\ddot{C}+\left( 3H+\frac{\dot{f}}{f}\right) \dot{C}= - V_C'(C)
\end{equation}
where 
 $V_C'(C) = dV/dC$.
 Using  $p_C \equiv \frac{1}{2} |\dot{C}|^2 - V_C$ and defining 
$\rho_C \equiv \frac{1}{2}|\dot{C}|^2 + V_C$,
  Eq.~(\ref{eq2})  can be recast as
 \begin{equation}
\dot{\rho}_C = -\left( 3H+\frac{\dot{f}}{f}\right) (\rho_C + p_C).
\end{equation}
For oscillations in a $V_C  \propto  C^4$ potential, $p_C = \rho_C/3$, so
$p_C = p_C^{(0)} (a^3 f)^{-4/3}$,
where $p_C^{(0)}$ is the initial value of the pressure.
The force in Eq.~(\ref{eq3}) then becomes
$-p_C^{(0)}f'(a^3 f)^{-4/3}  .$

As a specific example, 
consider the case $f(\Phi)= g(\Phi) = 1/\Phi =  {\rm exp} (-\lambda \phi)$.
This example assumes a single moduli field (the dilaton).
Later, we  will discuss the case of two or more moduli fields,
which is pertinent to perturbative string theory or 
non-perturbative M-theory \cite{HoravaW_D11SG_B,Witten,Lukas1,Lukas2}.
For $f(\Phi)= g(\Phi) = 1/\Phi$, an 
exponentially strong  force 
is induced by $p_C$  that 
adds  an effective potential to $V_{np}(\phi)$
equal to 
\begin{equation}
V_{eff}(\phi)=  \frac{3 p_C^{(0)}}{a^4} {\rm exp} (\lambda \phi/3).
\end{equation}
Note that $1/a^4 \propto T^4$, where the $T$ is the temperature of 
the radiation background.
$V_{eff}(\phi)$ is an exponentially increasing function that
provides a force pushing $\phi$ towards smaller values and
opposes $V_{np}$, which pushes $\phi$ toward higher values.  
Note that, expressed in terms of $\Phi$, the effective potential
is $V_{eff} \propto T^4 \Phi^{1/3}$.

Case (b), where $C$ is in thermal equilibrium, proceeds similarly.
Now the fluctuations in $C$  are non-negligible ($\nabla C \ne 0$)
and contribute to  the interaction  term 
$(f(\Phi) /2)|\partial C|^2 - g(\Phi) V_C$,
which does not obey the same simple relationship to the pressure 
$p_C$ as above. A different approach must be used to compute $V_{eff}$.
As above, we take a quartic potential $V_C = \epsilon C^4$
Under the assumption that $\Phi$ varies slowly compared to thermal 
interactions, we can transform $C \rightarrow \sqrt{f}C$ and
$g(\Phi) V_C = \epsilon g C^4 \rightarrow  (\epsilon g/f^2) C^4 \equiv \epsilon_{eff}
C^4$.   
In thermal equilibrium, the effective potential for a 
scalar field with quartic interactions is \cite{Kap} 
$V_{eff} = - (\pi^2 T^4 /30)[ 1 - (15/8) \epsilon_{eff} + \dots]$,
which includes a $\Phi$-dependent piece proportional to $(\pi^2 T^4/48)
(g/f^2)$. 
Whether this acts as an effective potential term that causes $\Phi$ to decrease
(stabilizes) or increase (destabilizes) depends critically on the dilaton
coupling to the kinetic energy.  For example, 
consider the case $f(\Phi) = g(\Phi) =1/\Phi$. Naively, based on the potential
energy term alone, $g(\Phi) ( \epsilon C^4)$,  one might suppose that the effective
potential is proportional to $g(\Phi) = 1/\Phi$, which is destabilizing.
However, when the kinetic energy contribution is properly included, 
\begin{equation}
V_{eff} = \frac{\pi^2}{48} \frac{T^4}{f(\Phi)} \propto 
T^4 \Phi = T^4  {\rm exp}(\lambda \phi).
\end{equation}
As in the case of coherent oscillations, $V_{eff}$ increases as $\Phi$ increases,
 which is the stabilizing condition we need.
In the remainder of the paper, we will consider   this 
case with thermal excitations, although the same considerations 
apply to the coherent oscillation case.

As shown in Fig.~1, the net effect is that $V_{eff}+ V_{np}$
at fixed temperature 
(dotted curves $V_{T_i}$)
has a temperature-dependent
minimum, $\Phi_{T_i}$,  about which the dilaton 
$\Phi$ oscillates.  The minimum lies at $\Phi_{T_i} < \Phi_{min}$.  As 
the universe expands and cools, the temperature decreases and
$V_{eff}$ decreases, as well.  The energy density at
$\Phi_{T_i}$  decreases and the value of $\Phi$ at the minimum
moves gradually towards $\Phi_{min}$.  

For this mechanism to work, an issue is that
oscillations in $\Phi$ about $\Phi_T$ must decay sufficiently
quickly that $\Phi$ does not jump over the barrier at low 
temperatures.  That is, even if $\Phi_T$ gently decreases towards
$\Phi_{min}$, it is conceivable that $\Phi$ is oscillating so
wildly about $\Phi_T$ that it is carried past the peak $\Phi_p$ 
at low temperatures when $V_{eff}(\Phi_p)  \le V_{np}(\Phi_p)$.
The large initial oscillations must be damped rapidly.
The greater is the damping rate, the larger can be
the initial oscillations,  and,  hence, the larger is 
the initial value of $\Phi$
that can be stabilized.

\begin{figure}
{\par\centering \resizebox*{4in}{4in}{\includegraphics{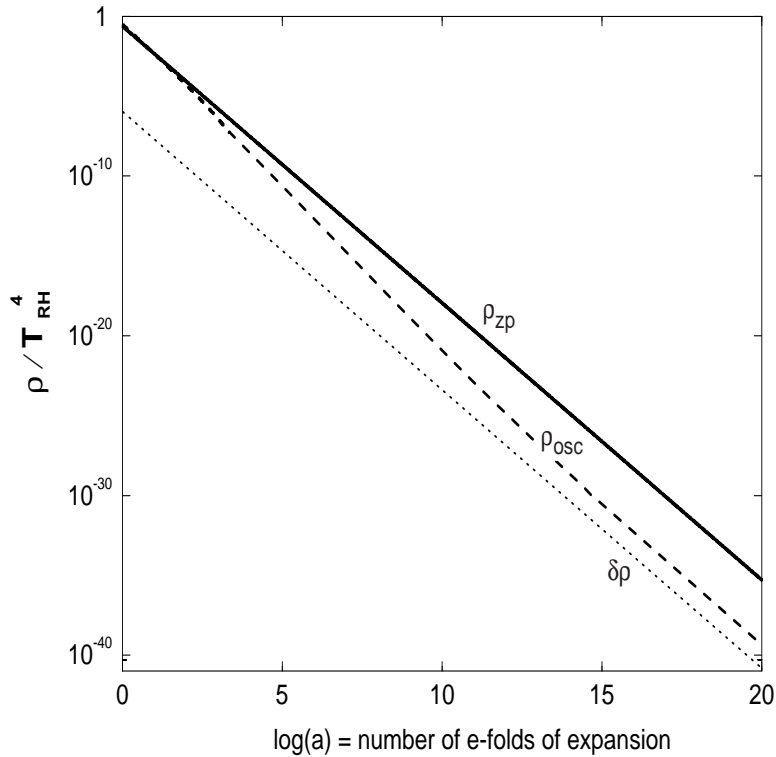}} \par}
\caption{\label{fig_Sr10Run} The evolution of the various energy densities
for the case of dilaton coupling $f = g =1/\Phi$.
 $T_{RH}$ is the reheat temperature after inflation.
The initial value of $\Phi$ was chosen to be $\Phi=10 \gg \Phi_{min}$. 
The figure shows how the
zero point ($\rho_{zp}$), oscillation ($\rho_{osc}$), and
 perturbation ($\delta\rho$) energy densities
evolve. In particular,
note that, although the system begins with $\rho_{osc} \sim \rho_{zp}$,
the oscillations are heavily damped 
after a few e-folds,
leading
to $\rho_{osc} \ll \rho_{zp}$. Furthermore, note that $\delta\rho$
(the contribution of inhomogeneity in all fields to the energy density) decays
at the same rate as $\rho_{zp}$, so inhomogeneity in the universe does not
come to dominate.}
\end{figure}

The total dilaton energy ($\rho_\phi$) at fixed temperature
can be split into the
zero-point energy ($\rho_{zp} \equiv V_{np}(\Phi_T) + V_{eff}(\Phi_T)$,
where $\Phi_T$ is the minimum of the finite temperature effective
potential)
and oscillation energy ($\rho_{osc} \equiv \rho_\phi - \rho_{zp}$). 
Thus for stabilization of the dilaton to be  robust, we need $\rho_{osc}$
to decay faster than $\rho_{zp}$. Figure~2 shows the
results of a numerical simulation for a typical case starting at
a temperature of approximately
$T_{RH}$ with $\Phi = 10 \gg \Phi_{min}$ and all of the components of the energy density
comparable. Note that initially $\rho_{osc} \sim \rho_{zp}$, but after $10$
e-folds of expansion it is about $4$ orders of magnitude smaller.
The relative damping of oscillation energy can be understood as follows:
the effective potential energy for $C$ decreases as $T^4$, like 
radiation.
 As $\Phi$ is
rolling along $V_{eff}$, 
the oscillation energy decays due to the
red shifting of its kinetic energy
{\it and} due to 
the fact that $V_{eff}$ decreases as the temperature decreases.
If $\Phi$ were frozen ($\dot{\Phi} = 0$)
at some value away from the minimum and all that happened is that
$V_{eff}$ decreases, 
the energy in the dilaton would decay
at the same rate as $V_{eff}$. With $\Phi$ oscillating ($\dot{\Phi} \ne 0$),
one has additionally the red shift of 
the dilaton kinetic energy; hence, $\rho_{osc}$ decreases more rapidly
than $V_{eff}$.
However, the rate of decay of the zero-point energy 
$\rho_{zp}$ is approximately
the same as $V_{eff}$. Thus,
 $\rho_{osc}$  decays faster than $\rho_{zp}$ and becomes negligible.
That is, the dilaton settles down near the minimum $\phi_T$ as the temperature
decreases.

A more rigorous argument shows that
 $\rho_{osc}$ decays faster
than $\rho_{zp}$ until $\rho_{osc}/\rho_\phi$ reaches a negligibly small value and
then the ratio remains roughly constant ($10^{-4}$ in Fig.~2).
The remaining oscillations are not important for our purposes since they are
too small to drive $\Phi$ past $\Phi_p$.
The decay rate of $\rho_{osc}/\rho_\phi$ is so rapid once oscillations
begin that it poses no significant constraint on our scenario. What does limit
the range of initial conditions is that, for sufficiently large $\Phi$,
there is insufficient time for oscillations to commence. We will return
to this point below when we determine how robust the stabilization
mechanism is.

Based on what has been learned from this example, it is 
straightforward to consider 
couplings different from $f(\Phi)=g(\Phi)= 1/\Phi$.
 A necessary (but insufficient) condition for the coupling to 
 produce a stabilizing $V_{eff}$ is that  $(g/f^2)' = d(g/f^2)/d\Phi >0$
 for the case of thermally excited $C$-fields. 
Hence, $f = g \propto 1/\Phi^n$ where $n>0$
 is a satisfactory form.
(Since $V_{eff}$ grows exponentially with  $\phi$ for all $n>0$,
the stabilization mechanism is not very sensitive to the power $n$.)

We have focused on the dilaton coupling $f(\Phi)$ to the kinetic
energy of the matter fields because they produce  a net, 
stabilizing, effective potential. We note that $S$ also couples
to the gauge fields via an interaction $h(\Phi) F_{\mu \nu} F^{\mu \nu}$,
where $ F_{\mu \nu} F^{\mu \nu} \approx B^2 - E^2$ in the case of
$U(1)$ gauge fields.  At high temperature, $<B^2> = <E^2>$, 
and so the gauge interaction adds zero effective potential for $\Phi$.
Hence, in the case of abelian gauge fields,
$h(\Phi) F_{\mu \nu} F^{\mu \nu}$ can be ignored for our purposes.

The dilaton coupling to the
axion is yet another interesting example. The kinetic energy of the 
axion couples to the dilaton with $f_A(\Phi)=1/2 \Phi^2$, a stabilizing form 
by the criterion outlined above.
However, 
the axion field is weakly coupled to matter,
and so it cannot be expected to be in thermal equilibrium with
the matter-fields. Instead, one can imagine that the axion has large
coherent time-variation, as discussed by Horne and Moore~\cite{ChaotCouplConst}.
This produces a steep,
stabilizing, effective 
potential $\propto \Phi^2 = {\rm exp}(2 \lambda \phi)$ which
forces  $\phi$  towards small values where it eventually
gets trapped in the  minimum of 
the combined potential due to the thermally excited
$C$-field and the non-perturbative potential $V_{np}$.
The  axion-induced 
force is not sustained for a very long time because
the strength is proportional to its pressure, $p_A \propto 1/ a^6$, 
which decays faster than the thermal energy.  However,
the brief contribution of the axion-induced force to dilaton
capture expands the
range of  $f(\Phi)$ and initial conditions for the dilaton
that are ultimately trapped.

How robust are the various stabilization mechanisms? That is,
beginning from initial conditions,  what is the probability
that $\Phi$ is trapped at $\Phi_{min}$?
A precise answer is not possible because there is no rigorous
understanding  of the initial  conditions.  We use 
plausible  estimates  
similar to Horne and Moore\cite{ChaotCouplConst} and others 
({\it e.g.}, we only consider
energy densities less than the Planck scale
and rough equipartition of kinetic and potential energies).
Originally, when the couplings between the dilaton and all
other fields were ignored, it appeared that a very narrow range
of initial conditions result in $\Phi$ being trapped at $\Phi_{min}$.
Formally, this is  a set of
measure zero if one imagines all possible initial values of $\Phi$ and
$\dot{\Phi}$ as being equally likely.
Barreiro {\it et al.}
propose a high-temperature thermal background of particles 
in order to increase the Hubble damping during the phase
when $\Phi$ evolves along the potential.
By increasing the damping of $\dot{\Phi}$,
this effect enhances the range of initial 
conditions by allowing $\Phi$ to lie somewhat further up the steep part of the
potential at $\Phi < \Phi_{min}$ and still not overshoot the peak at
$\Phi_p$. While this is an improvement, the range of allowed 
initial $\Phi$ remains  finite and narrow; formally, this is also a set of 
measure zero.  

Horne and Moore~\cite{ChaotCouplConst} argue that  
all possible values of $\Phi$
are not equally likely,
if couplings to the axion are properly included.
The nonlinear coupling between axion and dilaton causes 
the dilaton to follow a chaotic path of back and forth motion in
the potential in which large values of $\Phi>>\Phi_{min}$ are 
exponentially unlikely.
They argue that the effect can be taken into account by 
weighting the probability of $\Phi$ according to the K\"{a}hler
metric, which leads to a finite phase volume.
Fig.~3 shows two
representations of  the phase space of 
$\Phi$ and $A$. The horizontal 
bounding curves represent $A=0$ and $A=2 \pi m_{pl}/ \alpha $.  
 The  probability of  a given $\Phi'$ is proportional to the
 length of  the vertical segment joining  the upper and lower
 curves at $\Phi=\Phi'$.
Fig.~3a represents the naive expectation that 
 all combinations of initial $1 \le \Phi \le \infty$
and $0 \le A \le 2 \pi m_{pl}/\alpha$ are equally probable (all vertical 
segments joining the boundary have the same length).
In this case, the total volume is infinite.
 However, the non-linear coupling
between $\Phi$ and $A$ leads to chaotic dynamics at early times
which causes the probability distribution as a function of $\Phi$
to fall off as $1/\Phi^2$~\cite{ChaotCouplConst}.
Fig.~3b illustrates this distortion
of the phase space volume, which is now finite.
Horne and Moore conclude that, within the total volume,
the sub-volume of initial conditions that are ultimately
trapped at $\Phi_{min}$ is $\sim 14\%$ of the total volume,
corresponding to $\Phi$ near $\Phi_{min}$.
However, as later
pointed out by Banks {\it et al.}~\cite{ModulCosmo},  
the chaotic motion also causes
the evolution of unacceptably large inhomogeneities in the axion field.
In particular, the homogeneous component of the axion energy 
responsible for the chaotic motion decreases as $1/a^6$, whereas
the density inhomogeneities grow as $1/a^4$.  
So, while the universe may become trapped at $\Phi=\Phi_{min}$, the
density distribution is too inhomogeneous.

In judging  the stabilization mechanism proposed in this paper, we assume
the axion field is excited initially as well as the matter ($C$) fields. Hence,
we adopt the K\"{a}hler-weighted finite measure of the phase space
for initial $\phi$ as argued by Horne and Moore. 
To  estimate  what initial conditions are trapped, we impose
the conservative constraint
that our mechanism will rapidly stabilize the dilaton at $\Phi=\Phi_T$
beginning from some high initial temperature, {\it e.g.},
the reheat temperature after inflation, $T_{RH}$.
We determine the
maximum $\Phi$ for which
the dilaton completes one oscillation about $\Phi_T$ before the
temperature decreases to $10^{-3} T_{RH}$, say. After this
oscillation, $\rho_{osc}$ is already less than $\rho_{zp}$
and $\Phi$ is essentially caught near $\Phi_{T}$. We 
find that  $\Phi \le 50$ satisfies this
conservative condition, which 
encompasses $98\%$ of the initial phase space volume.
If we loosen our constraint by decreasing the bound
below $10^{-3} T_{RH}$, the fraction of allowed initial moduli space can
be made even closer to unity. 

\begin{figure}[t]
{\par\centering \resizebox*{4in}{4in}{\includegraphics{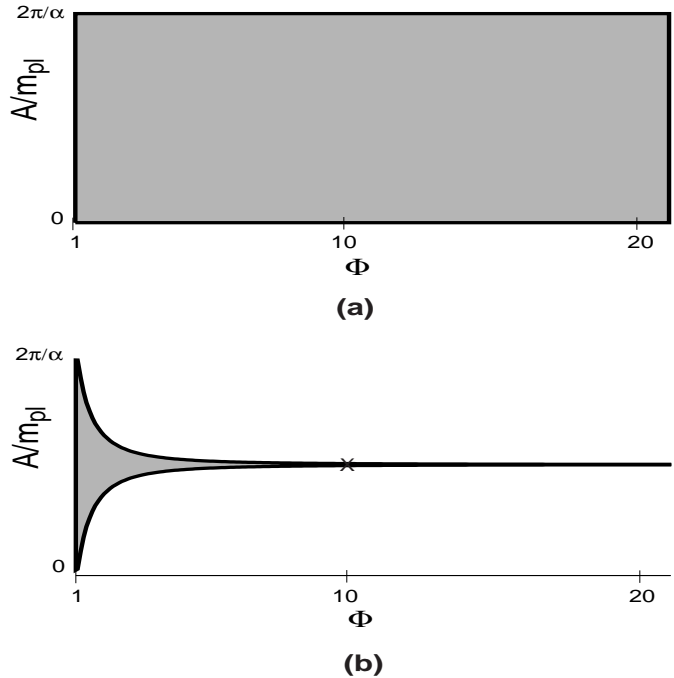}} \par}
\caption{\label{fig_ModSpaceVolume} A schematic illustration of initial
phase space volume. The relative likelihood of an initial $\Phi$ is represented
by  
the vertical distance between the curves bounding the shaded region.
 Naively, as shown in (a), all combinations of initial
$1 \le \Phi \le \infty$ and 
$0 \le A \le 2 \pi m_{pl} /\alpha$ might appear equally
probable, and the allowed volume of the shaded region is infinite.
However, based on the arguments of Horne and Moore,
the effective volume  of moduli space is defined by the K\"ahler
metric and is finite,  
as illustrated in (b).
The initial conditions used
in Fig.~2  are marked by ``X.''
}
\end{figure}

As an example, consider the case of
an initial value $\Phi=10$,
the case depicted in Fig.~2 and  marked by an ``X''
in  Fig.~3. 
This value lies outside the trapped region of Barriero {\it et al.},
which considers the Hubble damping effect, 
 and the trapped region of Horne and Moore, 
which considers only the dilaton-axion coupling. But this value
 lies well within the trapped region in our scenario, which 
includes the coupling between dilaton and $C$-field as well.
 Trapping all initial conditions
with $\Phi \le 10$ would be arguable progress if
Fig.~3a were 
correct, since this range would represent formally a set of measure
zero.  But, in Fig.~3b, this same range of initial conditions 
corresponds to $90\%$ of the total phase volume.

Figures~1 and 2 apply for
case of dilaton coupling $f(\Phi)=g(\Phi)= 1/\Phi$.
For a general $f(\Phi)$, we can ask what fraction of 
the K\"{a}hler-weighted volume of phase space for $\Phi$
is trapped at
$\Phi_{min}$.   Let us assume roughly equipartition
initial conditions in which the kinetic plus potential energy density in 
$\phi$ is comparable to the matter-field energy density.
For  $f(\Phi) = g(\Phi) = 1/ \Phi^n$, 
this implies an effective potential
$V_{eff} \sim \Phi^{n/3} \sim {\rm exp} (n \lambda \phi/3)$, which is
exponentially steep, 
sufficient to trap nearly $100\%$ of all initial conditions.

Unlike the case of Horne and Moore, our scenario does not suffer
from the problem of axion energy density inhomogeneities ($\delta\rho$).
In their scenario, energy density due to
inhomogeneities $\delta \rho$, which decays as $1/a^4$,
always  overtakes the homogeneous energy component,
the axion  kinetic energy, which decays as $1/a^6$.  
In our scenario, the homogeneous energy density is dominated by the
thermal energy  of the matter and gauge fields,  which decays 
as $1/a^4$.
(Here  $\delta\rho$ is defined
as the deviation in the $0-0$ component of the stress-energy tensor
due to perturbations in the dilaton, axion and $C$ fields as
well as the metric~\cite{Pert_MaBertshcinger}.)
Hence, as shown in Fig.~2, $\delta\rho$ decays at the same rate as
the total energy density ($\rho_{tot}$).  Assuming that the 
inhomogeneities are initially negligible, they remain negligible.

When two or more moduli fields exist, the situation becomes 
more complicated.  Both $f$ and $g$   take 
different forms.  
An example relevant to perturbative string theory or non-perturbative
M-theory \cite{HoravaW_D11SG_B,Witten,Lukas1,Lukas2} is
 $f[S,T] =  (3/Re[T]) + (\beta/Re[S])$ and $g[S,T] =
 1/(ST^3 f[S,T])$.
In models of the Ho\v{r}ava-Witten type,
  the dilaton $S$
 is replaced in the non-perturbative
 superpotential $W$ by
 \( S-\beta T \), where \( T \) is the orbifold modulus.
 Hence,  one can consider trapping in the $S- \beta T$
 direction; typically, an independent method  is needed to stabilize 
 the $S+ \beta T$ direction. If one supposes  a mechanism
 that fixes $Re[S+ \beta T] = \kappa$, where $\beta > 0$ and 
 $\kappa = {\cal O}(100) > 0$ 
 (as in the standard embedding), then  the effective 
 potential along the the $Re[S- \beta T]$ direction is similar
 to the  examples considered above. 
A technical difference is that, since the physical regime is 
$S>0$ and $T>0$,
the constraint,  $Re[S+\beta T]= \kappa$, prevents $\Phi= Re[S]$
from exceeding $\kappa$; so trapping is only required for $S \le \kappa 
= {\cal O}(100)$.
 The non-perturbative
 potential tends to push $\Phi = Re[S]$ to increase, but the 
thermal contribution due to the matter fields pulls
$\Phi$ back to smaller values.  
As in our toy model (see discussion of Eq.~(9)), the critical feature is that
the coupling to the kinetic energy produces a
a  stabilizing contribution to the thermal effective potential.
The trapping force  becomes  small at large $\Phi$.
 However, 
 an initial axion kinetic energy  
produces a steep, stabilizing potential at early times (until the
axion kinetic energy density becomes negligible compared to the dilaton energy).
 When all effects are included,
 the percentage of initial conditions that become trapped
rises to nearly 100\%, as before.

The lesson to be learned from this study goes beyond 
finding a long-sought  mechanism for stabilizing the dilaton.
What we have seen is that the cosmological background
can play an important role in the evolution and 
stabilization of moduli fields
and the determination of the present vacuum 
state. This is especially important for 
 nearly-flat, non-perturbative  
potentials with multiple vacua, as is common in supergravity and
superstring theories, where there is little guidance as to 
why one vacuum is observed and the others are irrelevant (at
least within our Hubble volume). 
A characteristic feature of these models is  non-linear 
sigma-model type couplings of the moduli fields to the kinetic energy of 
the matter  of the type considered here.   
Whereas  these couplings have been ignored in  past considerations
of the moduli problem, here we have seen that they can have a strong
influence in the early universe.
Hence, just as we have demonstrated
for the dilaton, we expect
the cosmological background 
 to have   significant effect on other moduli fields.

We thank M. Dine for  useful discussions..
The work was supported by the US Department of Energy grant
DE-FG02-91ER40671  (GH, PJS, DW) and DE-AC02-76-ER-03071 (BO).

\end{document}